# Low noise quantum frequency conversion of photons from a trapped barium ion to the telecom O-band

Uday Saha [1,2], James D. Siverns[1,2,3], John Hannegan[2,3], Qudsia Quraishi [4,2] and Edo Waks [1,2,3,5]

[1]Department of Electrical and Computer Engineering, University of Maryland College Park, MD, 20742.
[2]Institute for Research in Electronics and Applied Physics (IREAP), University of Maryland, College Park, MD, 20742.
[3]Department of Physics, University of Maryland College Park, MD, 20742.
[4]United States Army Research Laboratory, Adelphi, MD, 20783.
[5]Joint Quantum Institute (JQI), University of Maryland, College Park, MD, 20742.
[*]Corresponding Author: edowaks@umd.edu

## Abstract

Trapped ions are one of the leading candidates for scalable and long-distance quantum networks because of their long qubit coherence time, high fidelity single- and two-qubit gates, and their ability to generate photons entangled with the qubit state of the ion. One method for creating ion-photon entanglement is to exploit optically transitions from the $P_{1/2}$ to $S_{1/2}$ levels, which naturally emit spin-photon entangled states. But these optical transitions typically lie in the ultra-violet and visible wavelength regimes. These wavelengths exhibit significant fiber-optic propagation loss, thereby limiting the transfer of quantum information to tens of meters. Quantum frequency conversion is essential to convert these photons to telecom wavelengths so that they can propagate over long distances in fiber-based networks, as well as for compatibility with the vast number of telecom-based opto-electronic components. Here, we generate O-band telecom photons via a low noise quantum frequency conversion scheme from photons emitted from the $P_{1/2}$ to $S_{1/2}$ dipole transition of a trapped barium ion. We use a two-stage quantum frequency conversion scheme to achieve a frequency shift of 375.4 THz between the input visible photon and the output telecom photon achieving a conversion efficiency of 11%. We attain a signal-to-background ratio of over 100 for the converted O-band telecom photon with background noise less than 15 counts/sec. These results are an important step toward achieving trapped ion quantum networks over long distances for distributed quantum computing and quantum communication.

# Introduction

Quantum networks require optically active qubit systems with long coherence times. Trapped ions possess both of these properties, making them a promising candidate for scalable quantum networks.[1–4] They exhibit long qubit coherence times[5] and high-fidelity single- and two-qubits gates.[6–9] Additionally, we can readily manipulate them to produce single photons entangled with their internal spin state.[3,10–14] Trapped ions can be remotely entangled via an entanglement swap operation with their emitted photons. This method has been previously demonstrated with a node separation of few meters.[3,15] Recently, Krutyanskiy *et al.* have demonstrated remote ion-ion entanglement over 230 meters using the cavity-enhanced emission of near-infrared photons from the $P_{3/2}$ to $D_{5/2}$ transition.[4]

One approach to extend the ion-ion entanglement distance, is to produce telecom photons from ions utilizing quantum frequency conversion.[16] Several groups have reported the generation of telecom photons from ions using quantum frequency conversion of near-infrared photons from the $P_{3/2}$ to $D_{5/2}$ transition.[17–19] But these transitions have poor branching ratios which results in low emission probability, and may require cavities to boost their emission rates.[20,21] Ions emitting photons from the $P_{1/2}$ to $S_{1/2}$ dipole transition have a higher branching ratio and can provide direct entanglement of the emitted photon with the ground-state ion qubit.[3,10–14] However, this transition is often in the ultraviolet or visible wavelengths which are particularly difficult to convert to the telecom bands because of the large wavelength separation between the input and target telecom photon. Previously, we reported the generation of C-band telecom photons from a trapped barium ion from the $P_{1/2}$ to $S_{1/2}$ dipole transition.[22] However, the reported signal-to-background ratio of the C-band photon was only 0.04 because of the pump laser-induced Raman anti-Stokes noise photons, limiting its practical usage in quantum networks.[22]

In this work, we report a low noise quantum frequency conversion scheme to generate O-band 1287 nm single photons using the photons emitted from the $P_{1/2}$ to $S_{1/2}$ dipole transition of a barium ($^{138}$Ba$^+$) ion. We use two cascaded quantum frequency conversion stages to avoid spontaneous parametric down-conversion (SPDC) noise photons that would be present in single-stage 493 nm to O-band frequency conversion. In each stage of quantum frequency conversion, we choose the separation between the wavelength of the pump laser and the output photon to be large enough to reduce the impact of Raman anti-Stokes noise photons. We measure a signal-to-background ratio of over 100 for the O-band single photon, which is limited by the background counts of the detector and the probability of producing and collecting a 493 nm photon. Additionally, we achieve an overall photon conversion efficiency of 11% through our two-stage quantum conversion scheme. By measuring the second-order intensity correlation function[23], we verify that this two-stage quantum frequency conversion preserves photon-number quantum statistics of visible photons produced from the barium ion. These results make scalable long-distance quantum networks with ultraviolet and visible trapped ion transitions possible using existing telecom infrastructure and photonic integrated circuits.

# Experimental Details and Methods

Figure 1 shows a block diagram illustration of the quantum frequency conversion setup. This setup uses a two-step quantum frequency conversion process to convert 493 nm photons to 1287 nm, which is within the telecom O-band. Each stage implements difference frequency generation in a zinc-doped periodically-poled lithium niobate (PPLN) waveguide. In the first stage, we convert the 493 nm photons to 781 nm using a 1342 nm pump laser. The conversion process uses a Sagnac interferometer-like configuration to convert both orthogonal polarizations of the photon.[24] This stage was characterized and recently used for ion-photon entanglement in a previous work, with a reported conversion efficiency of 35% for each orthogonal polarization mode.[24] Light from the first stage is coupled to a single-mode fiber, which is then combined with the second-stage pump laser using free-space optics. We then directly inject these combined signals into the second PPLN waveguide using an aspheric lens.

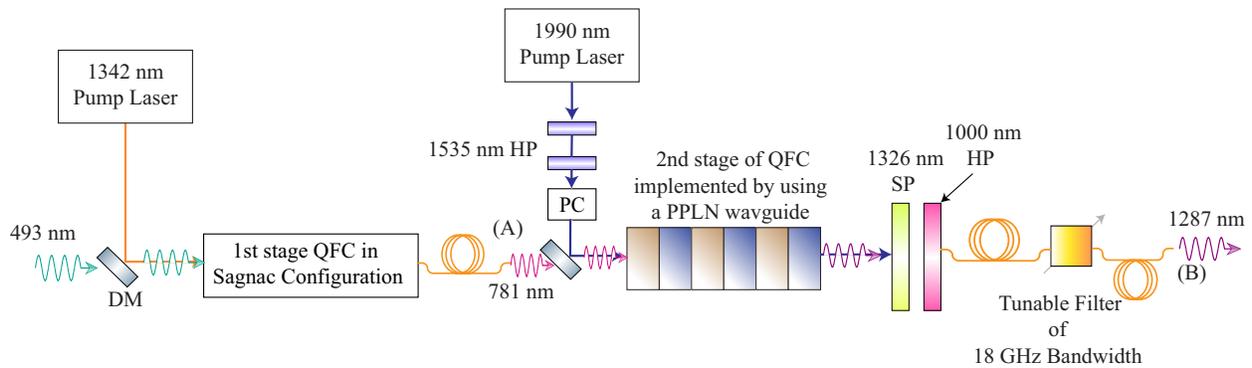

**Fig. 1** A schematic representation of the two-stage quantum frequency conversion to generate O-band single photons from a trapped barium ion. Here QFC: quantum frequency conversion, HP: high pass filter, SP: short pass filter, DM: dichroic mirror, PC: polarization control, PPLN: periodically poled lithium niobate. The photons from the first stage are coupled to a single-mode fiber and injected into the second stage of QFC using free space optics. The output O-band photons from the second stage are coupled to a single-mode fiber and routed through the fiber-coupled tunable fiber of 18 GHz bandwidth.

We convert the 781 nm single photons to 1287 nm single photons in the second stage converter using a 1990 nm pump laser. This stage was not operated in a Sagnac-like configuration like the first stage and, as a consequence, only converts one of the orthogonal polarizations of the photon.

To eliminate noise photons arising from the second-stage pump laser at the target wavelength of the output photons, we utilize two 1535 nm high pass filters (Semrock: FF01-1535/LP-25) in the optical path of the 1990 nm pump laser before going to the second PPLN waveguide. After the second-stage frequency conversion, we use a 1326 nm short-pass optical filter (Semrock: FF01-1326/SP-25) to block the second-stage pump laser and transmit the target 1287 nm photons. In addition, we utilize a 1000 nm long-pass optical filter (Thorlabs Inc: FEL1000) to block unconverted 781 nm photons and noise photons coming from the first stage frequency conversion setup. Finally, we use a fiber-coupled tunable filter with a bandwidth of 18 GHz (WL Photonics Inc: WLTF-NM-P-1280-40/0.10-SM-0.9/1.0-FC/APC) to reduce noise photons coming from the second stage frequency conversion. At the operating target wavelength of 1287 nm, the transmission through the tunable filter is around 69%.

## Results and Analysis

We first characterize the conversion efficiency of the second stage conversion system using a linearly-polarized 781 nm laser. Fig. 2(a) shows the photon conversion efficiency of the second stage from the input fiber to the output of the tunable filter (from point (A) to point (B) in Fig. 1). We fit the data points in Fig. 2(a) to $\eta = \eta_0 sin^2(\frac{\pi}{2}\sqrt{\frac{P}{P_m}})$, where $\eta$ is the conversion efficiency at pump power $P$.[22] The parameter $P_m$ is the pump power that achieves the peak conversion efficiency, denoted as $\eta_0$. From the fit, we determine that the peak conversion efficiency is $\eta_0$ =35.6%, achieved at a pump power of $P_m$=278 mW (Fig. 2(a)). Combined with the ~35% efficiency of the first stage, we achieve a total conversion efficiency of 11% through both stages, which includes losses in connectors and optical filters.

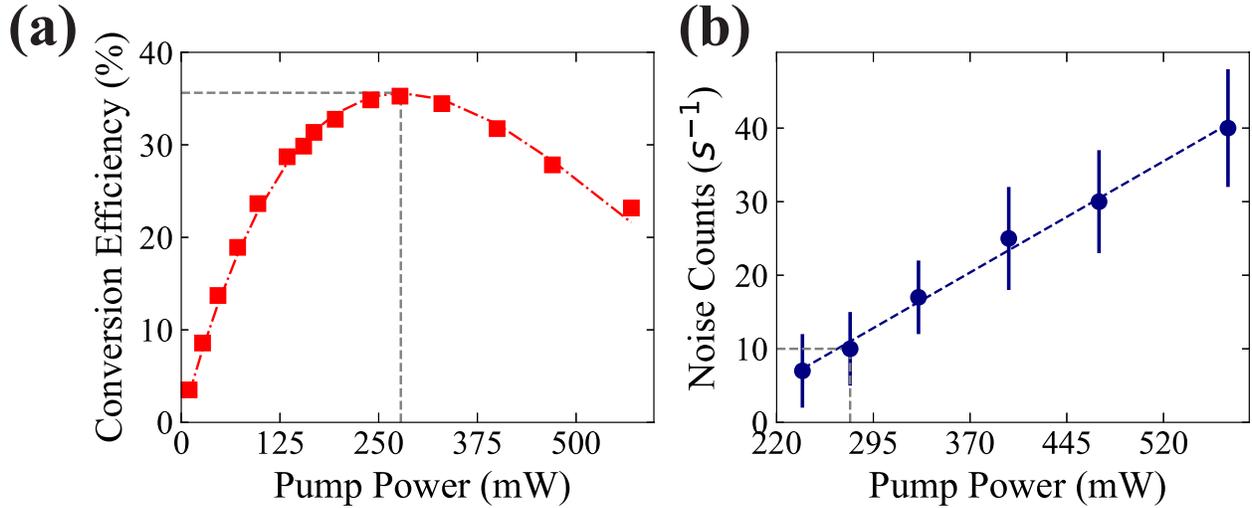

Fig. 2 (a) The photon conversion efficiency of the second stage as a function of the second stage pump power. (b) total noise count as a function of the second stage pump power, where the first stage pump power was set to around 180 mW. The gray dashed lines indicate the characteristics of the operating point that we use in our experiment. The error bars in panel (a) were obtained from a two-minute average for each data point and account for the drift in optical alignment and power in the pump laser. The error bars are plotted but not visible on this scale. The blue line in (b) is a linear fit to the data. The error bars in (b) include the shot noise limited accuracy along with the count fluctuations due to the light leakage.

Next, we analyze the noise of the two-stage conversion system as a function of second stage pump power while operating the first stage pump at around 180 mW. We then obtain the noise counts by measuring the signal at 1287 nm at the output of the second stage. Figure 2(b) plots the noise count rate of the two-stage frequency conversion scheme as a function of the second-stage pump power. We detect the background noise using a superconducting nanowire single photon detector (SNSPD) with a reported detection efficiency of 87% at 1287 nm and a dark count rate of $60\pm10$ s$^{-1}$. To extract the noise contribution from the quantum frequency conversion alone, we subtract the noise count rate of the detectors from the overall background rate. At peak conversion efficiency, we achieve a noise count rate of only $10\pm5$ s$^{-1}$, which is significantly lower than the detector dark count rate. The linear dependence of the noise count rate with pump power indicates that the dominant noise source is Raman anti-Stokes scattering.[25] We note that this low dark count rate is achieved with a filtering bandwidth of 18 GHz. We could significantly reduce this noise count rate even further by using a narrower bandwidth filter such as an etalon or high-finesse cavity.

In order to convert single photons from a trapped ion, we use the experimental layout illustrated in Figure 3 (a). We use a single $^{138}$Ba$^+$ as the photon source. The experimental setup to trap and collect light from the ion is described in detail in Ref. [24,26]. We collect the emitted 493 nm photons from both the front and back reentrant windows of the vacuum chamber housing the ion trap. We send the photons collected from the back window to a photomultiplier tube (PMT). We collect and fiber-couple the photons emitted out of the front side of the chamber using a 0.6 numerical aperture objective lens (Photon Gear 15920-S_B) and direct them to the two stage quantum frequency conversion setup. We select the pump power in each quantum frequency conversion stage to achieve a maximum photon conversion efficiency. We note that the barium ion emits an equal superposition of two orthogonally polarized modes[26] and the second stage of quantum frequency conversion can only convert one of these modes. Thus, we suffer an additional factor of two loss, resulting in an efficiency of 5.25%, for converting visible photons from the barium ion. This loss can be recovered by extending the second-stage conversion system to convert both polarizations using either a Sagnac configuration[18,24] or by using two orthogonally polarized crystals.[19] We detect the generated O-band photons using an SNSPD. To match the polarization of the O-band photons to the preferred polarization axis of the SNSPD, we use a free-space quarter and half waveplates before fiber coupling into the detector. Both 493 nm photons, detected on the PMT, and O-band photons, detected on the SNSPD, are time-tagged with a timing resolution of 80 ps (limited by the SNSPD timing jitter) relative to a synchronization pulse for measuring the second-order intensity correlation function.

To measure the signal-to-background ratio of frequency-converted O-band photons, we plot the time-resolved O-band photon counts without background subtraction in Fig. 3(b) relative to the synchronization pulse. We calculate the total counts in a 41.6 ns photon detection window, indicated by green dashed lines in Fig. 3 (b). This window captures 95% of the total O-band photon counts. We determine the noise counts by summing the photons detected during a window of the same size but delayed by 300 ns relative to the peak of the O-band photon. This delay is sufficient such that no photon counts arrive from the ion, but we still collect noise counts. In Fig. 3(b), we indicate this noise window by the vertical red solid lines. We then subtract these noise counts from the total counts in the O-band photon and divide the resultant quantity by the noise counts to determine the signal-to-background ratio.

We obtain a signal-to-background ratio of $110\pm1.8$ in the O-band photon for this detection window (the shaded green area in Fig. 3(b)). This value is over 2750 times larger than that reported in our previous work which converted photons from a barium ion to the telecom C-band.[22] We can improve the signal-to-background at the expense of the overall signal-rate by optimizing the size of the photon window. For example, if we choose a 20.8 ns photon window, equivalent to 50% of the total O-band photon counts, we get a signal-to-background ratio of $165.4\pm3.9$. For the rest of the analysis in this article, we consider the 41.6 ns photon window that captures most of the counts (95%) in the O-band photon. The signal-to-background ratio reported here is limited by detector background counts and the collection efficiency of 493 nm photons. We can reduce the dark counts by a factor of 5 by using an SNSPD with a lower dark count rate of 2 counts/sec.[19] The collection efficiency could also be improved by incorporating photon collection optics inside the vacuum chamber of the trap.[20,27] The probability of detecting a photon within this window is 2.6×10$^{-4}$, measured by averaging the photon count rate over the experimental runtime. Over the course of the experiment, we detected a total of *3.92 × 10$^5$* and *3.59 × 10$^3$* O-band photon and noise counts, respectively.

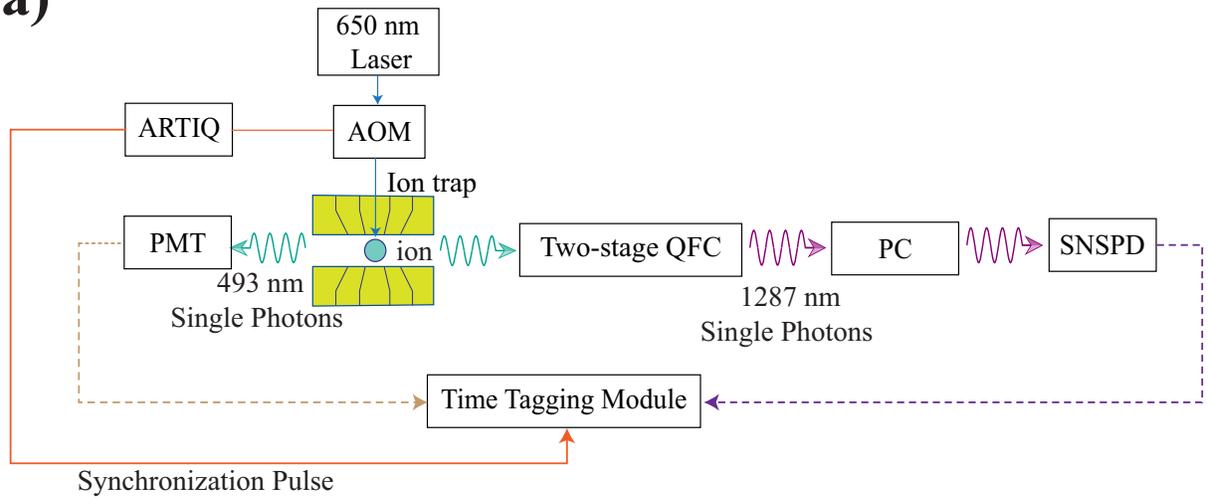
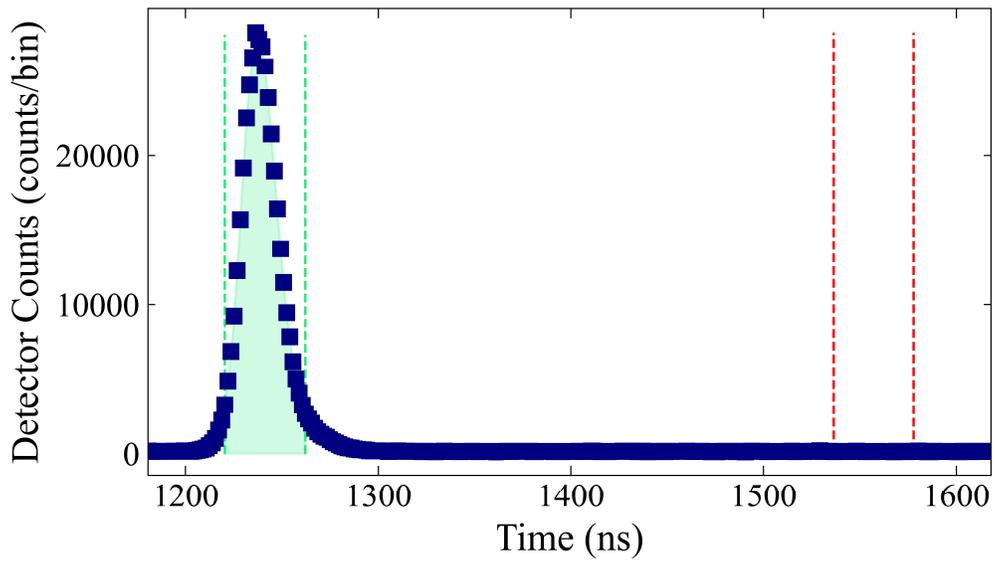
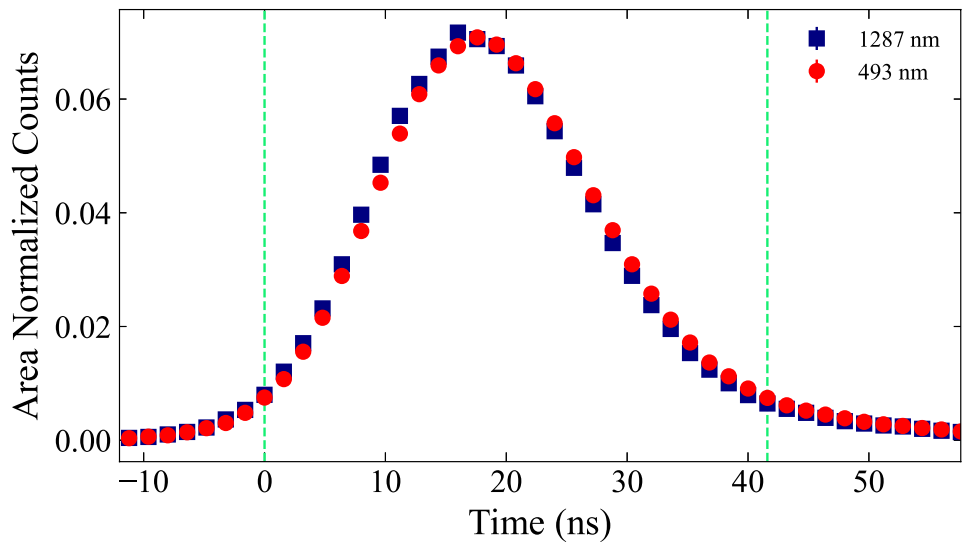

Fig. 3 (a) The experimental layout of measuring the time-resolved fluorescence and second-order correlation function of 493 nm and 1287 nm photons where ARTIQ is the control sequence system, AOM is an acousto-optic modulator and PC is personal computer. (b) Time-resolved counts of O-band 1287 nm photons without background subtraction. We indicate the noise window by vertical solid red lines. (c) The temporal pulse shapes of 493 nm and 1287 nm photons. We perform background subtraction and area-normalization in (c). In both (b) and (c), the green dashed vertical lines indicate the photon window that we consider, and the green shaded area is the photon area in (b). We plot the data points with shot noise limited error bars in both (b) and (c), but the error bars are not visible at this scale.

We compare the temporal pulse shape of the 493 nm and O-band photon in Fig. 3(c) in order to demonstrate that quantum frequency conversion preserves the temporal profile of the pulse. For each temporal pulse shape, we perform background subtraction and normalize each curve by its total area. From the figure we see complete overlap of the temporal pulse shape of the 493 nm signal with the 1287 nm signal, which demonstrates the preservation of the temporal pulse shape after quantum frequency conversion.

To verify that quantum frequency conversion preserves the photon-number quantum statistics, we measure the second-order intensity correlation between 493 nm photons emitted by the ion detected on the PMT and the converted O-band photons detected on the SNSPD (Fig. 3(a)). Fig. 4 shows the normalized second-order correlation function $g^{(2)}(n)$, where $n$ is the number of the experimental cycles between photon-detection events. We obtain $g^{(2)}(0)=0.04\pm0.01$ without background subtraction, verifying that the two-stage conversion process preserves the single photon characteristics of the ion emission. The deviation from perfect anti-bunching can be fully explained by the background level due to detector dark counts and injected noise from the quantum frequency conversion. Using the measured values of the noise and dark counts, we calculate the theoretical $g^{(2)}(0)$ to be 0.05 which within one standard deviation of the experimental value.[22] (The detailed calculation of the value $g^{(2)}(0)$ can be found in the Supporting Information.)

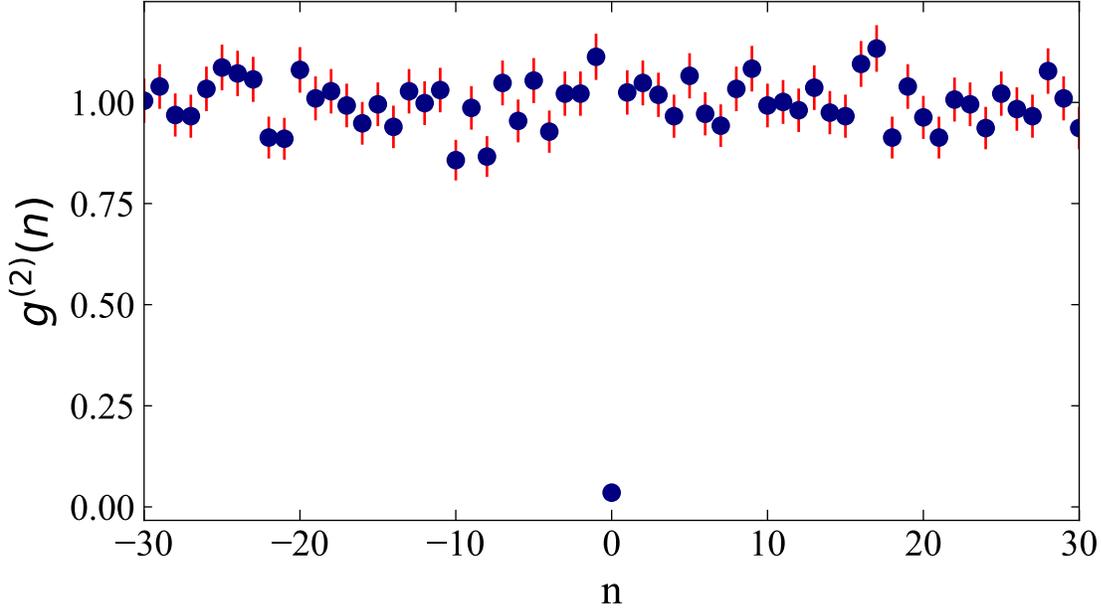

**Fig. 4** The second-order correlation function between the 493 nm and O-band photon. Here $n$ represents the number of experimental cycles between photon-detection events. We measure $g^{(2)}(0)=0.04\pm0.01$ verifying the quantum nature of the generated O-band photon after the two-stage quantum frequency conversion.

# Discussion

In summary, we generate O-band telecom 1287 nm photons from a trapped ion using 493 nm photons emitted from an $P_{1/2}$ to $S_{1/2}$ dipole transition. We use a two-stage quantum frequency conversion scheme to ensure low background noise counts and demonstrate preservation of their temporal and quantum properties. By carefully choosing the wavelength of the pump laser combined with optical filtering in each stage of frequency conversion, we attain a low-noise broad frequency shift with 11% overall photon conversion efficiency. Although the detector background counts limit the signal-to-background ratio, we achieve a ~2750 times improvement in signal-to-background ratio from our previous reported results.[22] The frequency conversion scheme presented in this work provides a clear pathway for attaining broad frequency shifts in the down-conversion of photons with low noise. These results enable trapped ion systems to use the existing telecom infrastructure to distribute entanglement over long distances.[1,18] Additionally, conversion to telecom opens the possibility to use foundry-fabricated telecom photonic integrated circuits to create a scalable and programmable network of trapped ion quantum computers.[1,28,29] Our results represent an important step towards interconnecting trapped ions and distributing entanglement over long-stance for scalable quantum networks.

# Acknowledgments

**Funding**
We would like to acknowledge the support from the National Science Foundation (Grant Nos. OIA2134891, EFMA1741651 and OIA2040695) and the Air Force Office of Scientific Research (Grant Nos. FA95501610421 and FA95501810161).

**Author contributions:** U. Saha and E. Waks conceived the idea. U. Saha, J. D. Siverns and J. Hannegan performed the experiment, and U. Saha analyzed the data. All authors contributed to write, edit, and review the manuscript.

**Competing interests:** All authors declare no conflict of financial interest.

All part numbers and company references are given for technical purposes, and their mention does not represent an endorsement on the part of the U.S. government. Other equivalent or better options may be available.

**Data and materials availability:** All the data and codes that support the finding of this study are available from corresponding author upon request.

# References

(1) Monroe, C.; Kim, J. Scaling the Ion Trap Quantum Processor. *Science* **2013**, *339* (6124), 1164–1169.
(2) Duan, L.-M.; Monroe, C. Colloquium: Quantum Networks with Trapped Ions. *Rev. Mod. Phys.* **2010**, *82* (2), 1209–1224.
(3) Stephenson, L. J.; Nadlinger, D. P.; Nichol, B. C.; An, S.; Drmota, P.; Ballance, T. G.; Thirumalai, K.; Goodwin, J. F.; Lucas, D. M.; Ballance, C. J. High-Rate, High-Fidelity Entanglement of Qubits Across an Elementary Quantum Network. *Phys. Rev. Lett.* **2020**, *124* (11), 110501.
(4) Krutyanskiy, V.; Galli, M.; Krcmarsky, V.; Baier, S.; Fioretto, D. A.; Pu, Y.; Mazloom, A.; Sekatski, P.; Canteri, M.; Teller, M.; Schupp, J.; Bate, J.; Meraner, M.; Sangouard, N.; Lanyon, B. P.; Northup, T. E. Entanglement of Trapped-Ion Qubits Separated by 230 Meters. *Physical Review Letters* **2023**, *130*, 050803.
(5) Wang, P.; Luan, C.-Y.; Qiao, M.; Um, M.; Zhang, J.; Wang, Y.; Yuan, X.; Gu, M.; Zhang, J.; Kim, K. Single Ion Qubit with Estimated Coherence Time Exceeding One Hour. *Nat. Commun.* **2021**, *12* (1), 233.
(6) Harty, T. P.; Allcock, D. T. C.; Ballance, C. J.; Guidoni, L.; Janacek, H. A.; Linke, N. M.; Stacey, D. N.; Lucas, D. M. High-Fidelity Preparation, Gates, Memory, and Readout of a Trapped-Ion Quantum Bit. *Phys. Rev. Lett.* **2014**, *113* (22), 220501.
(7) Ballance, C. J.; Harty, T. P.; Linke, N. M.; Sepiol, M. A.; Lucas, D. M. High-Fidelity Quantum Logic Gates Using Trapped-Ion Hyperfine Qubits. *Phys. Rev. Lett.* **2016**, *117* (6), 060504.
(8) Srinivas, R.; Burd, S. C.; Knaack, H. M.; Sutherland, R. T.; Kwiatkowski, A.; Glancy, S.;

Knill, E.; Wineland, D. J.; Leibfried, D.; Wilson, A. C.; Allcock, D. T. C.; Slichter, D. H. High-Fidelity Laser-Free Universal Control of Trapped Ion Qubits. *Nature* **2021**, *597* (7875), 209–213.

(9) Clark, C. R.; Tinkey, H. N.; Sawyer, B. C.; Meier, A. M.; Burkhardt, K. A.; Seck, C. M.; Shappert, C. M.; Guise, N. D.; Volin, C. E.; Fallek, S. D.; Hayden, H. T.; Rellergert, W. G.; Brown, K. R. High-Fidelity Bell-State Preparation with $^{40}Ca^{+}$ Optical Qubits. *Phys. Rev. Lett.* **2021**, *127* (13), 130505.

(10) Crocker, C.; Lichtman, M.; Sosnova, K.; Carter, A.; Scarano, S.; Monroe, C. High Purity Single Photons Entangled with an Atomic Qubit. *Opt. Express* **2019**, *27* (20), 28143–28149.

(11) Auchter, C.; Chou, C.-K.; Noel, T. W.; Blinov, B. B. Ion–photon Entanglement and Bell Inequality Violation with $^{138}Ba^{+}$. *J. Opt. Soc. Am. B* **2014**, *31* (7), 1568.

(12) Moehring, D. L.; Maunz, P.; Olmschenk, S.; Younge, K. C.; Matsukevich, D. N.; Duan, L.-M.; Monroe, C. Entanglement of Single-Atom Quantum Bits at a Distance. *Nature* **2007**, *449* (7158), 68–71.

(13) Kobel, P.; Breyer, M.; Köhl, M. Deterministic Spin-Photon Entanglement from a Trapped Ion in a Fiber Fabry–Perot Cavity. *npj Quantum Information* **2021**, *7* (1), 1–7.

(14) Blinov, B. B.; Moehring, D. L.; Duan, L.-M.; Monroe, C. Observation of Entanglement between a Single Trapped Atom and a Single Photon. *Nature* **2004**, *428* (6979), 153–157.

(15) Maunz, P.; Moehring, D. L.; Olmschenk, S.; Younge, K. C.; Matsukevich, D. N.; Monroe, C. Quantum Interference of Photon Pairs from Two Remote Trapped Atomic Ions. *Nat. Phys.* **2007**, *3* (8), 538–541.

(16) Kumar, P. Quantum Frequency Conversion. *Opt. Lett.* **1990**, *15* (24), 1476–1478.

(17) Walker, T.; Miyanishi, K.; Ikuta, R.; Takahashi, H.; Vartabi Kashanian, S.; Tsujimoto, Y.; Hayasaka, K.; Yamamoto, T.; Imoto, N.; Keller, M. Long-Distance Single Photon Transmission from a Trapped Ion via Quantum Frequency Conversion. *Phys. Rev. Lett.* **2018**, *120* (20), 203601.

(18) Bock, M.; Eich, P.; Kucera, S.; Kreis, M.; Lenhard, A.; Becher, C.; Eschner, J. High-Fidelity Entanglement between a Trapped Ion and a Telecom Photon via Quantum Frequency Conversion. *Nat. Commun.* **2018**, *9* (1), 1998.

(19) Krutyanskiy, V.; Meraner, M.; Schupp, J.; Krcmarsky, V.; Hainzer, H.; Lanyon, B. P. Light-Matter Entanglement over 50 Km of Optical Fibre. *npj Quantum Information* **2019**, *5* (1), 1–5.

(20) Christoforou, C.; Pignot, C.; Kassa, E.; Takahashi, H.; Keller, M. Enhanced Ion–cavity Coupling through Cavity Cooling in the Strong Coupling Regime. *Sci. Rep.* **2020**, *10* (1), 15693.

(21) Casabone, B.; Friebe, K.; Brandstätter, B.; Schüppert, K.; Blatt, R.; Northup, T. E. Enhanced Quantum Interface with Collective Ion-Cavity Coupling. *Phys. Rev. Lett.* **2015**, *114* (2), 023602.

(22) Hannegan, J.; Saha, U.; Siverns, J. D.; Cassell, J.; Waks, E.; Quraishi, Q. C-Band Single Photons from a Trapped Ion via Two-Stage Frequency Conversion. *Appl. Phys. Lett.* **2021**, *119* (8), 084001.

(23) Scully, M. O.; Zubairy, M. S. *Quantum Optics*; Cambridge University Press, 1997.

(24) Hannegan, J.; Siverns, J. D.; Quraishi, Q. Entanglement between a Trapped-Ion Qubit and a 780-Nm Photon via Quantum Frequency Conversion. *Phys. Rev. A* **2022**, *106* (4), 042441.

(25) Pelc, J. S.; Ma, L.; Phillips, C. R.; Zhang, Q.; Langrock, C.; Slattery, O.; Tang, X.; Fejer, M. M. Long-Wavelength-Pumped Upconversion Single-Photon Detector at 1550 Nm:

Performance and Noise Analysis. *Opt. Express* **2011**, *19* (22), 21445–21456.
(26) Siverns, J. D.; Li, X.; Quraishi, Q. Ion-Photon Entanglement and Quantum Frequency Conversion with Trapped Ba+ Ions. *Applied Optics* **2017**, *56* (3), B222–B230.
(27) Shu, G.; Chou, C.-K.; Kurz, N.; Dietrich, M. R.; Blinov, B. B. Efficient Fluorescence Collection and Ion Imaging with the "tack" Ion Trap. *J. Opt. Soc. Am. B* **2011**, *28* (12), 2865.
(28) Saha, U.; Siverns, J. D.; Hannegan, J.; Prabhu, M.; Quraishi, Q.; Englund, D.; Waks, E. Routing Single Photons from a Trapped Ion Using a Photonic Integrated Circuit. *Physical Review Applied* **2023**, *19* (3), 034001.
(29) Saha, U.; Waks, E. Design of an Integrated Bell-State Analyzer on a Thin-Film Lithium Niobate Platform. *IEEE Photonics J.* **2022**, *14* (1), 1–9.

# Supporting Information
# Low noise quantum frequency conversion of photons from a trapped barium ion to the telecom O-band

Uday Saha [1,2], James D. Siverns[1,2,3], John Hannegan[2,3], Qudsia Quraishi [4,2] and Edo Waks [1,2,3,5]

[1]Department of Electrical and Computer Engineering, University of Maryland College Park, MD, 20742.
[2]Institute for Research in Electronics and Applied Physics (IREAP), University of Maryland, College Park, MD, 20742.
[3]Department of Physics, University of Maryland College Park, MD, 20742.
[4]United States Army Research Laboratory, Adelphi, MD, 20783.
[5]Joint Quantum Institute (JQI), University of Maryland, College Park, MD, 20742.
[*]Corresponding Author: edowaks@umd.edu

## Single photon production from a trapped barium ion

We Doppler-cool a trapped barium ($^{138}$Ba$^+$) ion for 100 μs using all polarizations of 650 nm and 493 nm light before we perform a sequence of 500 photon production attempts. Each photon production attempt lasts for 10 μs and consists of 8 μs of optical pumping by 493 nm $\pi$ polarized light and 650 nm $\pi$ and $\sigma-$ polarized light to prepare the ion in the $|5D_{3/2}, m_j=+3/2\rangle$ state followed by a 600 ns delay with no light incident on the ion to ensure that the turn-off latency of the acousto-optical modulators (AOMs) do not impact the photon excitation [1]. Following this delay time, we send a trigger pulse lasting for 200 ns to the time-tagging module to act as a reference for any photon detection event. After that, we use a 200 ns pulse of 650 nm $\sigma+$ polarized light to excite the ion to the $|6P_{1/2}, m_j=+1/2\rangle$ state from where it can spontaneously decay into the $|6S_{1/2}, m_j= \pm1/2\rangle$ manifold, emitting a single 493 nm photon. The repetition rate of photon production attempts is 100 kHz. Fig. S1 represents the timing sequence of 493 nm photon production.

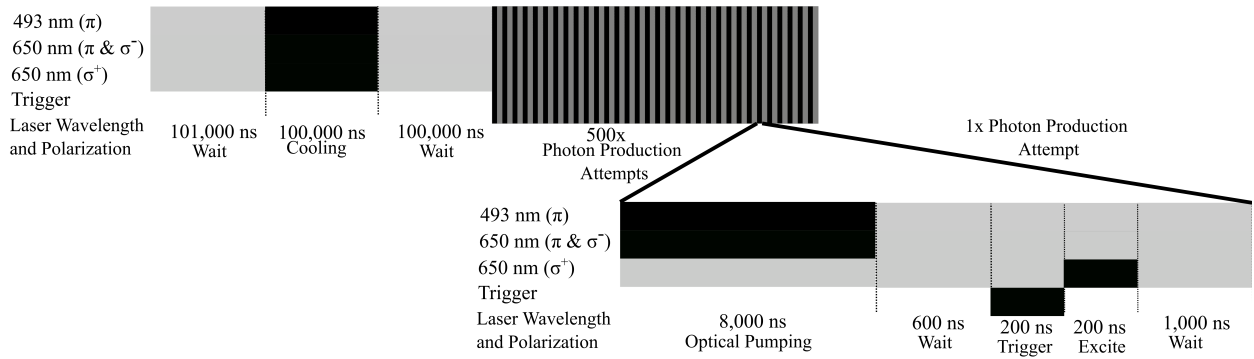

**Fig. S1** The experimental timing sequence of producing 493 nm single photons from a trapped barium ion ($^{138}$Ba$^+$).

## Time-resolved counts of 493 nm photon

Fig. S2 represents the time-resolved 493 nm photon counts detected on the photomultiplier tube without background subtraction. Like O-band photon counts, we calculate the total counts in a 41.6 ns photon detection window, indicated by green dashed lines in Fig. S2. We indicate the noise detection window by red solid vertical lines in Fig. S2. Over the course of the experiment, we measured a total count of $1.27 \times 10^6$ and $5.49 \times 10^4$ in the visible-photon and noise-detection windows, respectively. The signal-to-background ratio of detecting 493 nm visible photon is $23.1 \pm 0.1$ limited by the background counts of the photomultiplier tube ($857 \pm 30$ counts/sec).

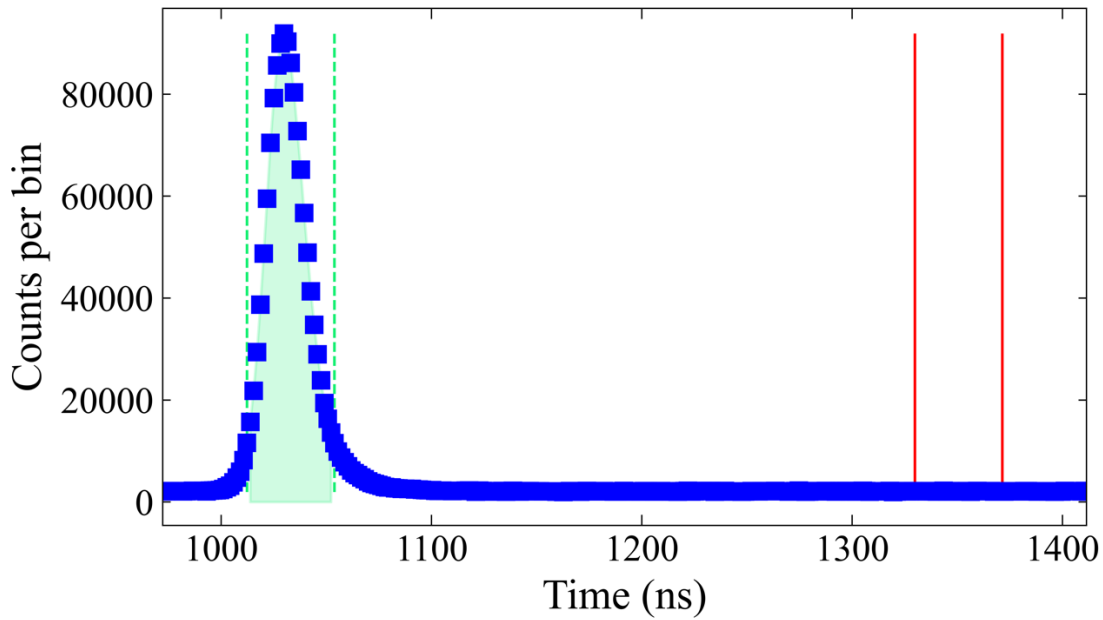

**Fig. S2** Time-resolved counts of visible 493 nm photon without background subtraction. The green dashed and red solid vertical lines indicate the photon and noise detection windows, respectively.

The green shaded area is the photon area of the visible photon. We plot the data points with error bars in the figure which correspond to shot noise-limited accuracy. However, the error bars are very small to be separated from the data points.

## Calculation of expected $g^{(2)}(0)$

To calculate the expected $g^{(2)}(0)$, we first calculate the unnormalized second order correlation ($G^{(2)}(0)$) between O-band photons and 493 nm photons by the following equation.

$$G^{(2)}(0) = \frac{C_1^S C_2^N + C_1^N C_2^S + 2\, C_1^N C_2^N}{R}$$

Here $C_1^S$ and $C_1^N$ represents the background subtracted O-band photon counts and corresponding noise counts respectively, detected by the SNSPD. Similarly, $C_2^S$ and $C_2^N$ are the background subtracted 493 nm photon counts and corresponding noise counts respectively detected by the photomultiplier tube. R represents the total number of photon production attempts. To find $g^{(2)}(0)$, we use the equation, $g^{(2)}(0) = \frac{G^{(2)}(0)}{a}$ where we define the normalization factor, $a = \frac{C_1^{total} C_2^{total}}{R}$. Here, $C_{1(2)}^{total}$ is the total number of counts detected in the O-band (493 nm) photon window including the noise photons. In our experiment, we measured $C_1^S = 3.92 \times 10^5$, $C_1^N = 3.59 \times 10^3$, $C_2^S = 1.27 \times 10^6$ and $C_2^N = 5.49 \times 10^4$. Over 4.27 hours of experimental run time, we measured a total photon production attempt of $R = 1.54 \times 10^9$.

## References

(1) Siverns, J. D.; Li, X.; Quraishi, Q. Ion-Photon Entanglement and Quantum Frequency Conversion with Trapped Ba+ Ions. *Applied Optics* **2017**, *56* (3), B222–B230.